\documentclass[10pt]{article}
\usepackage[OE]{express}
\usepackage[]{graphicx}
\usepackage{bm}
\usepackage{textcomp,enumitem,array}
\usepackage{amsmath}
\usepackage[mathscr]{euscript} 

\begin{document}
\title{A differential optical shadow sensor for sub-nanometer displacement measurement and its application to drag-free satellites}
\author{Andreas Zoellner,\authormark{1,2} Si Tan,\authormark{2,3,*} Shailendhar Saraf,\authormark{2,5} Abdul Alfauwaz, \authormark{2} Dan DeBra,\authormark{1} Sasha Buchman,\authormark{2}and John A. Lipa\authormark{2,4}}

\address{\authormark{1}Department of Aeronautics and Astronautics, 476 Lomita Mall, Stanford University, CA, 94305, USA\\
\authormark{2}Hansen Experimental Physics Laboratory, 452 Lomita Mall, Stanford University, CA, 94305, USA\\
\authormark{3}Department of Applied Physics, 348 Via Pueblo, Stanford University, CA, 94305, USA\\
\authormark{4}Department of Physics, 382 Via Pueblo Mall, Stanford University, CA, 94305, USA\\
\authormark{5}SN\&N Electronics, Inc., San Jose, CA, 95125, USA}

\email{\authormark{*}stan1987@stanford.edu} 


\begin{abstract}
We present a method for 3D sub-nanometer displacement measurement using a set of differential optical shadow sensors. It is based on using pairs of collimated beams on opposite sides of an object that are partially blocked by it. Applied to a sphere, our 3-axis sensor module consists of 8 parallel beam-detector sets for redundancy. The sphere blocks half of each beam's power in the nominal centered position, and any displacement can be measured by the differential optical power changes amongst the pairs of detectors. We have experimentally demonstrated a displacement sensitivity of 0.87 $\rm{nm/\sqrt{Hz}}$ at 1 Hz and 0.39 $\rm{nm/\sqrt{Hz}}$ at 10 Hz. We describe the application of the module to the inertial sensor of a drag-free satellite, which can potentially be used for navigation, geodesy and fundamental science experiments as well as ground based applications.
\end{abstract}

\ocis{(120.6085) Space instrumentation; (350.6090) Space optics; (130.6010) Sensors.} 

\bibliographystyle{osajnl}

\section{Introduction}
The recently reported detection of gravitational radiation from colliding black holes \cite{abbott2016observation} has renewed interest in the possibilities for observing similar very low frequency events using detectors in space \cite{audley2017laser}. One aspect of mission design is the use of drag-free spacecraft to reduce orbit perturbations that affect the signal/noise ratio. Drag-free technology also has more immediate application to measurements of the Earth's geoid \cite{nguyen2015three,Canuto2008DragFree}. Over the last few years we have developed a novel optical technique to simplify the instrumentation with the aim of reducing cost and improving mission reliability. We report on the first implementation of the optics for a residual drag sensor that can have a performance that meets the requirements for a range of space missions and ground-based uses. 

\subsection{Drag-free satellites}
A drag-free satellite is one that is freed from all forces other than gravity \cite{Lange1964DragFree}. Uses range from navigation \cite{Department1974Satellite} to geodesy \cite{Canuto2008DragFree} to fundamental physics \cite{debra_drag-free_2003, Everitt2011Gravity, buchman2013lisa}. The detection of very low frequency gravitational waves in space is probably the most challenging application for drag-free satellites, while geodesy and navigation have much less stringent requirements. 

At the heart of a drag-free satellite is a Gravitational Reference Sensor (GRS) that measures the satellite's displacement relative to a geodesic, i.e. an orbit that is only influenced by gravity. A GRS consists of a free floating test mass (TM) that acts as the reference and a system to measure its position relative to the vehicle. Ideally this TM is shielded against all non-gravitational forces such as residual gas drag, solar radiation pressure, and electro-magnetic forces. In the  GRS being developed at Stanford \cite{sun_advanced_2005}, an important component is the Differential Optical Shadow Sensor (DOSS): a sub-nanometer displacement sensor that measures the relative motion between the satellite to which the GRS is mounted and the free floating TM. The error signals from the GRS are then used to control thrusters that drive them to null.

The shape of the TM has some implications for the displacement measurement. Spherical TMs have been used in the TRIAD I and Gravity Probe B missions \cite{Everitt2011Gravity,doi:10.1063/1.57412}, while cubic TMs are used in Laser Interferometer Space Antenna (LISA) Pathfinder (LPF) \cite{PhysRevLett.116.231101} and Gravity field and steady-state Ocean Circulation Explorer (GOCE) \cite{Canuto2008DragFree}. In general, spherical TMs have the advantage of a less complex control system as only the translational degrees of freedom need to be controlled \cite{Gerardi2014Invited}, while disadvantages include the need to spin the sphere or fully map its surface in order to handle geometric imperfections and reflectivity variations over the surface for interferometric readouts  \cite{cavalleri2009increased}. The sensor described in this paper is designed for a spherical TM.


\subsection{Displacement measurement}
There are a number of different technologies that can be used to measure the displacement between two objects. For this paper we primarily focus on technologies relevant for drag-free satellites and accelerometers, i.e. displacement measurements without physical contact to the objects. There are two common measurement technologies: capacitive and optical. The former was used on the first drag-free satellite, TRIAD I as well as Gravity Probe B. The LISA Pathfinder satellite used both capacitive and optical (interferometric) measurements.

\subsubsection{Capacitive displacement measurement}
The capacitance $C$ between two objects is, to first order, inversely proportional to the distance $d$ between the objects. For the simplified case of two large parallel plates with area $A$ and small $d$, the relationship is $C \propto A/d$. This shows a fundamental trade-off of capacitive sensing: The displacement sensitivity $\delta d/d \propto \delta C/C$ decreases with increasing gap size $d$, while the DOSS sensitivity is independent of $d$. However, large gap size is advantageous for a drag-free applications as disturbance forces generally decrease with the gap size. For optical sensing of the TM only, the main gap-dependent disturbances are the Brownian force noise from molecular impacts in a constrained volume \cite{cavalleri2009increased} $(\propto 1/d^2)$ and the electrical stiffness induced by the interaction of image charges $(\propto 1/d)$, DC voltages $(\propto 1/d^2)$ and patch fields $(\propto 1/d^3)$ from the TM with those on the housing \cite{doi:10.1063/1.57412}. If capacitive sensing and electrostatic forcing of the TM is used then the acceleration noise induced by sensor back-action and the applied control forces dominates \cite{doi:10.1063/1.57412, Gerardi2014Invited}. It is caused by dielectric losses and by the interaction of the applied electric fields with the image charges, DC voltages and patch fields \cite{sun_advanced_2005,PhysRevLett.116.231101} on the TM and housing; where all these effects scale as $1/d$ \cite{doi:10.1063/1.57412}. 


\subsubsection{Optical displacement measurement}
The two primary optical displacement measurement technologies are: interferometric and intensity based measurements. 

Interferometric measurements are an integral part of any gravitational wave detector because of their inherently high sensitivity. However they are rarely used for drag free control primarily because of complexity, except when an interferometric measurement is available from  other experimental requirements. 

The state of the art interferometric measurements for a satellite based system with a cubical TM is the sensor on board of LISA Pathfinder with a performance of $35\:\rm{fm/\sqrt{Hz}}$ at $\geq 60$ mHz \cite{Heinzel2005Successful,PhysRevLett.116.231101}. This system measures a single degree of freedom over a short path length between two cubic TMs inside the satellite. The drag free control is implemented with capacitive sensors except for the x axis of TM one.

The sensor described in this paper is an intensity based sensor. Optical position sensors were first used in follow up missions to TRIAD, 1974 and later, (see for example \cite{S2,S3} and references therein). In 2003 Leitner proposed an optical position sensor for a spherical TM: "a simple optical readout system using photodiode and LED arrays would be employed to provide symmetrical illumination and proof mass position detection" with a resolution of 0.1-0.5 mm \cite{leitner2003investigation}. In 2005 Acernese et al \cite{acernese2005optical} proposed an optical readout system for the right rectangular prism TM of LISA; to be used as a back-up redundant system to its primary capacitive readout, while the Stanford team started the development of an interferometer based optical sensor meeting LISA requirements \cite{sun_advanced_2005,allen2009optical}. The development of a medium sensitivity (nanometer rather than picometer range) Differential Optical Shadow Sensor (DOSS), optimized as the reference sensor for drag-free control or as the primary TM read-out for geodesy and autonomous navigation applications, was initiated at Stanford in 2007 and has been upgraded since that time \cite{sun2008differential,sun2009modular,zoellner2013integrated}. The results described here were obtained with greatly reduced beam power and well-collimated beams. A recent study showed an estimation and control scheme using a differential optical shadow sensor for three-axis drag-free control of a satellite with a single thruster \cite{nguyen2015three}. An ultra-precise one-dimensional optical shadow sensor was developed by Lockerbie \cite{lockerbie2011first,lockerbie2014violin} for the measurement of the violin modes of the quartz fibers suspending the Laser Interferometer Ground Observatory (LIGO) mirrors.

Our design has the capability of operating with very large gaps between the TM and its housing, greatly reducing the impact of internal parasitic forces. It also allows the controlled return of the TM from any location in its housing to the centered position in the event of a temporary loss of control. The components are now readily available in space qualified form.

\subsection{Paper overview}
In this paper, we present a sub-nanometer displacement measurement method using a Differential Optical Shadow Sensor and show its application to drag-free satellites. We start with an introduction to DOSS in section \ref{section:DOSS}. The experimental setup is described in section \ref{setup}. The noise analysis is presented in section \ref{theory}. A discussion of the experimental results is presented in section \ref{results}, followed by conclusions in section \ref{conclusion}.

\section{The differential optical shadow sensor}
\label{section:DOSS}
The basic operating principle of the differential optical shadow sensor was first described for application to drag-free satellites by Lange \cite{Lange1964DragFree}. To our knowledge this concept has not previously been implemented in hardware. Some early results with one-dimensional measurements have been reported \cite{zoellner2013integrated}. 

In our device eight light beams are arranged around a spherical TM (radius = 12.7 mm) such that the TM blocks half of each beam when it is in its nominal centered position with respect to the housing. The light beams are fixed in the housing and the light intensity reaching the opposite side of the housing is measured. Any relative movement between the TM and the housing leads to a change of intensity measured by two or more detectors.

The beams originate from a single fiber coupled laser diode light source that is split into equal parts with fiber optic beam splitters. A differential measurement between two beams is obtained by routing the received intensities to a fiber-coupled balanced photodetector. Four of these differential measurements are taken sequentially at high speed to measure the displacement of the TM in all three directions, while providing redundancy in the z direction. 

\begin{figure}[htbp!]
  \centering
  \includegraphics[width=0.4\textwidth]{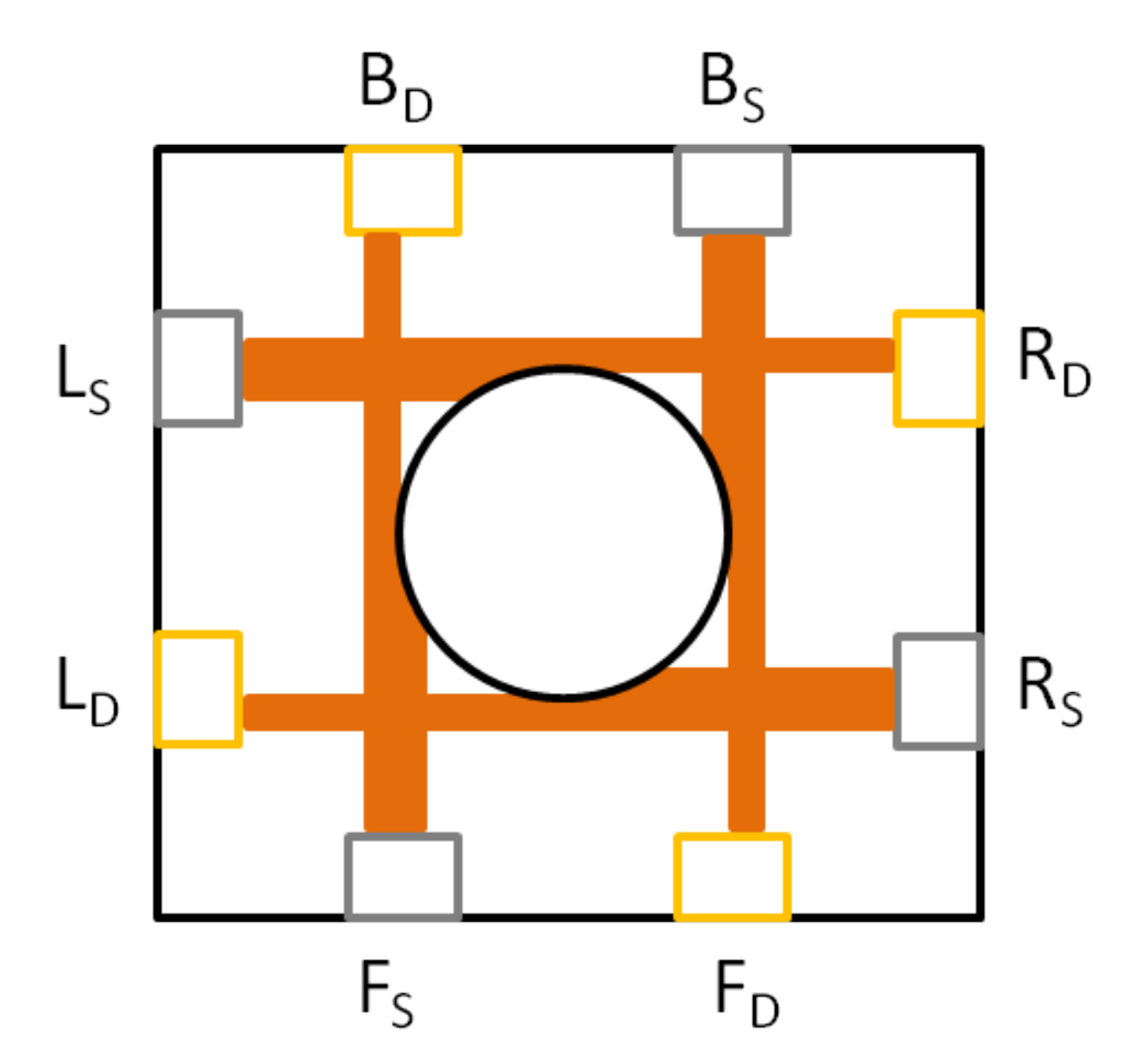}
  \includegraphics[width=0.44\textwidth]{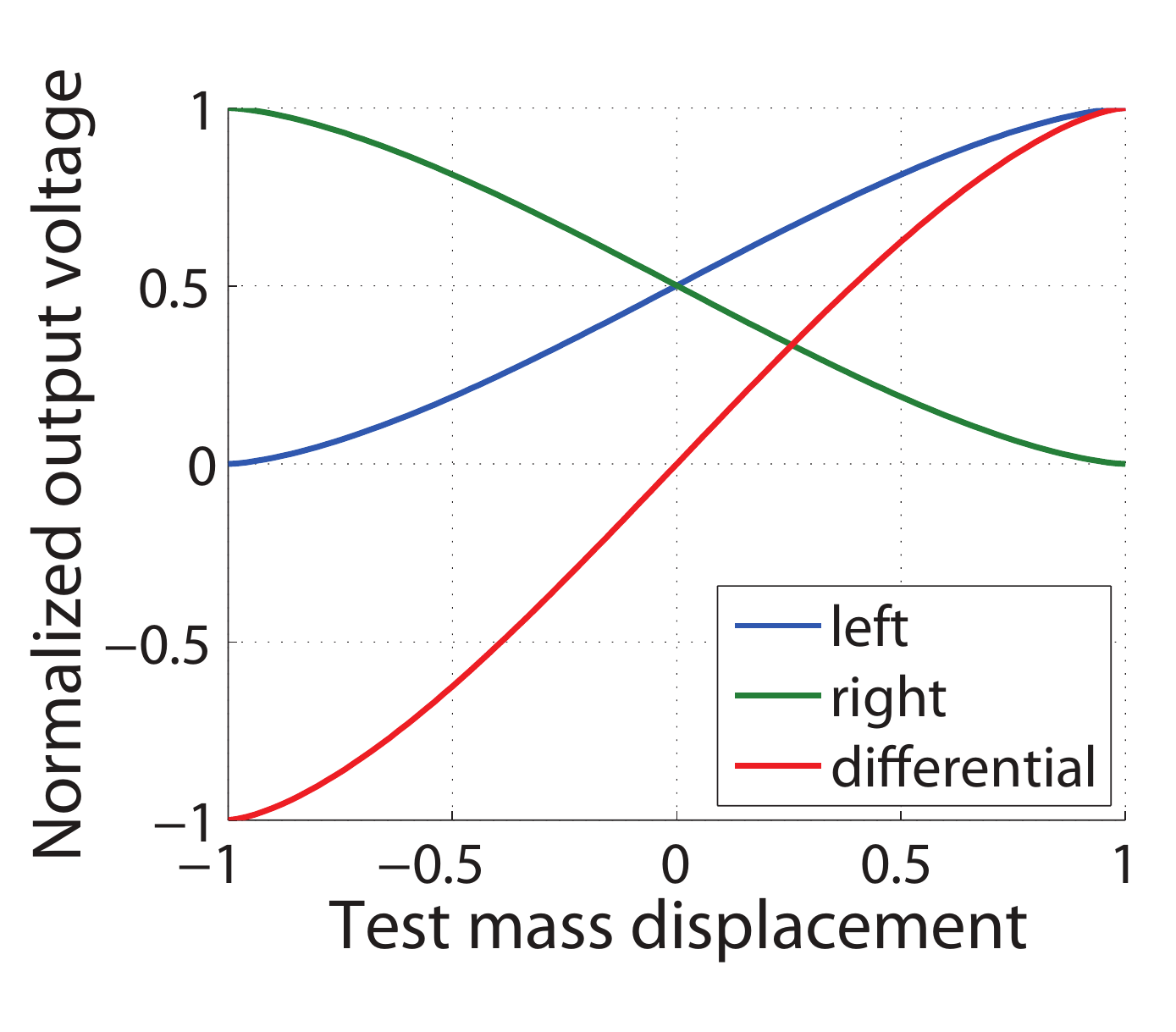}
  \caption{Left: principle of the DOSS for 2D measurements, detectors (indicated by subscript D) are shown in light orange, light sources (indicated by subscript S) are in grey, the letters indicate position left, right, back and front. Right: Normalized signals from a pair of detectors vs normalized TM displacement.}
  \label{fig:DOSS-Principle}
\end{figure}

In order to minimize the disturbance of the TM due to photon pressure variations, the beams are arranged such that the sum of the forces and the moments exerted on the TM by the photons are zero when the TM is centered. This is illustrated in the 2D schematic in Fig. \ref{fig:DOSS-Principle}. The four beams shown in this figure have no in-plane net force on the TM. There is a second layer of four beams that cancels out the moments and the vertical forces. This geometry was chosen to divide the signals into 3 orthogonal axes to provide signals for the spacecraft thrusters.

The range of the sensor is approximately 2.7 mm, which is the nominal diameter of the slightly convergent DOSS beams. Since in operation the relative motion is $\sim$ few $\rm{\mu m}$, there is no need to account for nonlinearities at the range boundaries. Furthermore, since the radius of the beams is much smaller than that of the TM, the differential optical signals are linear to a few percent for TM translations of up to 0.1 mm. Due to the use of a large gap between the housing and the TM an issue arises regarding the feasibility of initial centering of the TM within the DOSS beams. We note that when the TM is not in the operational DOSS range, its position with respect to the 8 quadrants of the housing is uniquely determined by the combination of obscured, partially obscured, and not obscured DOSS beams. Thus, to acquire the TM after launch we plan to make use of the 3-axis micro-Newton thrusters that would be on the spacecraft to move the beams until they are affected by the TM. The velocities involved could be $\sim$ 1 $\rm{\mu m/s}$. We have simulated 2D capture successfully and note that as more beams are crossed, the location of the TM will become easier to compute, allowing convergence in all 3 directions. We have also provided a video showing the simulation of a 2D capture in the supplementary document.
\section{Experimental setup}
\label{setup}
\begin{figure}[htbp!]
  \centering
  \includegraphics[width=0.35\textwidth]{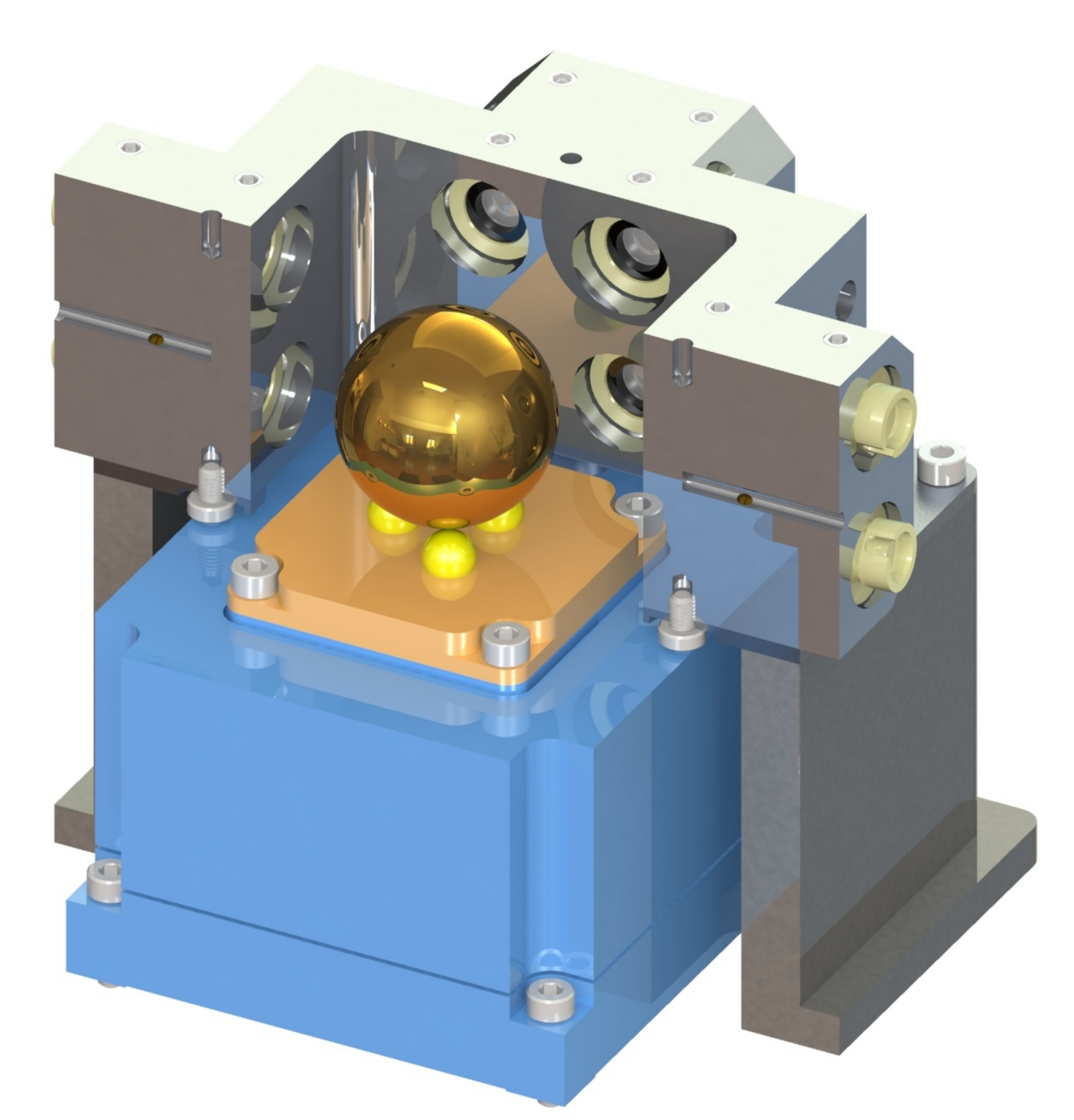}\hspace{1cm}
  \includegraphics[width=0.37\textwidth]{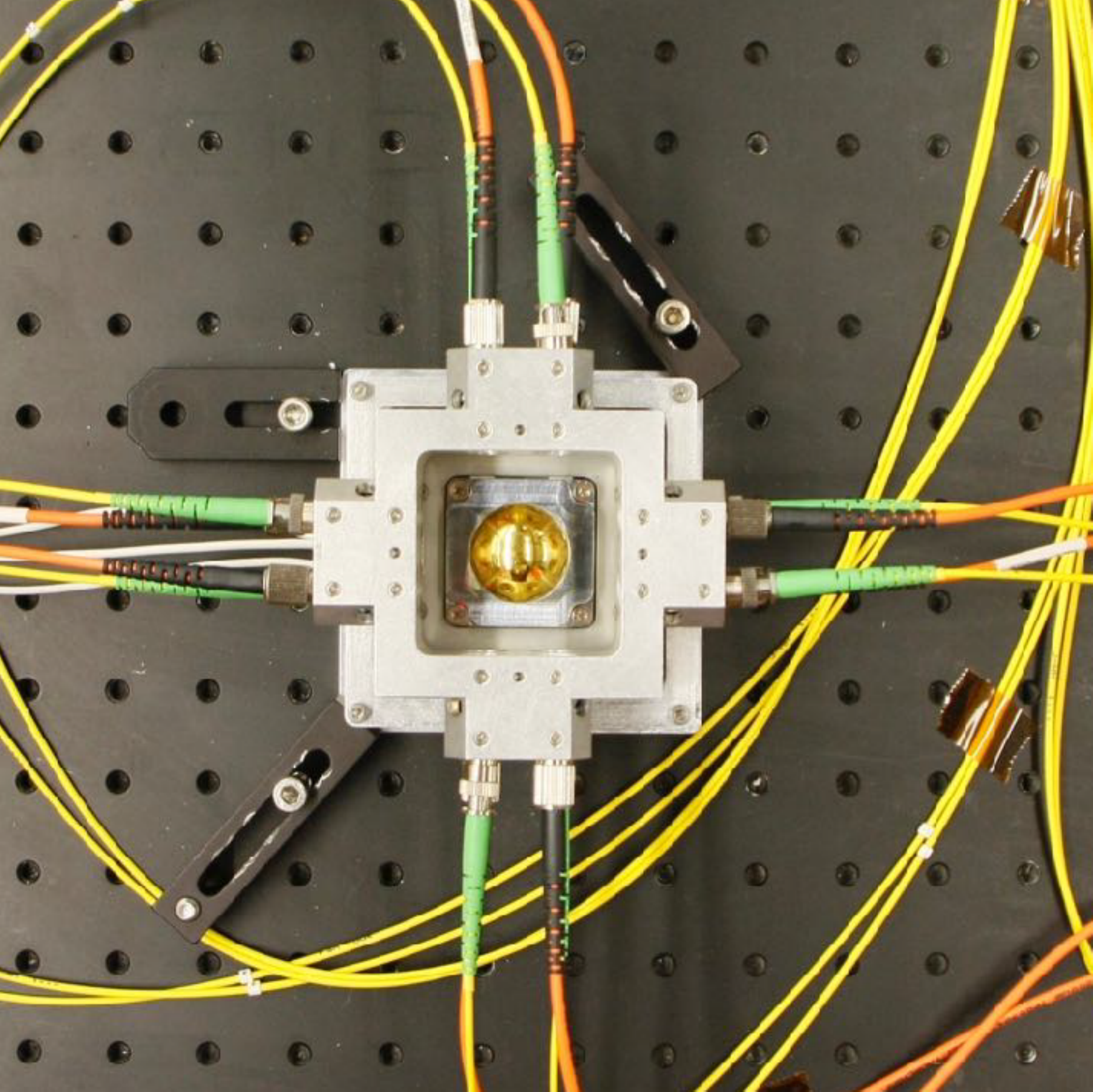}
  \caption{(a). Rendering of DOSS setup with cut through the housing (grey);kinematic TM support (yellow); 3-axis piezo actuated stage (blue); (b). Photo of the device top view. Input beams are carried by the yellow SM fibers, output beams are collected by orange MM fibers.}
  \label{fig:DOSS_setup}
\end{figure}

The test setup is illustrated in Fig. \ref{fig:DOSS_setup}. In this laboratory configuration, the TM is kinematically mounted on a 3 axis piezo actuated stage, a 
Nano-M350 model from Mad City Labs with a USB20-3 piezo controller. The Nano-M350 has a range of motion of 50 $\rm{\mu m}$ in the horizontal (X and Y) directions and 20 $\rm{\mu m}$ in the vertical (Z) direction. Its precision is 0.1nm and 0.04nm in the horizontal and vertical directions respectively. It also has an internally generated calibration signal that can be added to the displacement of any axis. The piezo stage and the housing around the TM are mounted on an optical table for seismic isolation. The 8 light beams are coupled into and out of the housing through a set of 16 identical 
Schafter \& Kirchhoff (60FC-4-M15-26) collimators, with 15 mm focal length and 420-700 nm spectral range, each attached to an input or output fiber. The fibers from the light source to the collimators are single mode (SM) which allows higher spatial intensity stability than with multi-mode (MM) fibers. Between the collimators and the photodetectors MM fibers are used for easier beam alignment relative to SM fibers.

The optical source was a fiber coupled laser from Thorlabs 
(Model S1FC660) with a wavelength of 660nm and a spectral linewidth of $<1$ nm (FWHM). Seven 50/50 beam splitters (1+2+4) from 
Schafter \& Kirchhoff (FBS-635-X-50/50-5xAPC-100-100) provided the 8 equal intensity beam sources for the DOSS. The housing design and recess configuration of the collimator reduces the coupling of the scattered light. 

The Thorlabs 
PDB450A-AC series Si detectors were chosen for their fiber coupled inputs, switchable gain, and separate output channels for single/differential measurements. These detectors have two inputs and three outputs. Each of the two inputs is connected to a monitor output port and the differential signal is amplified and connected to the third output port. Their responsivity is 0.53 A/W and the noise equivalent power is 3.3 $\rm{pW/\sqrt{Hz}}$ for a 4 MHz bandwidth. The translational and rotational alignment of the collimators perpendicular to the beam can be adjusted with set-screws.

The fiber optical design was chosen to keep all active components such as laser diodes and photodetectors well away from the innermost thermal enclosure of the sensor. We note that the DOSS does not occupy the top and bottom faces of the housing. These faces will be used for UV-LED based charge management \cite{saraf2016ground}, and a caging mechanism \cite{Zoellner2013Caging}. An additional interferometric displacement sensor required for state of the art gravitational wave detection would be accommodated in a corner of the housing.

\begin{figure}[htbp!]
  \centering
  \includegraphics[width=0.75\textwidth]{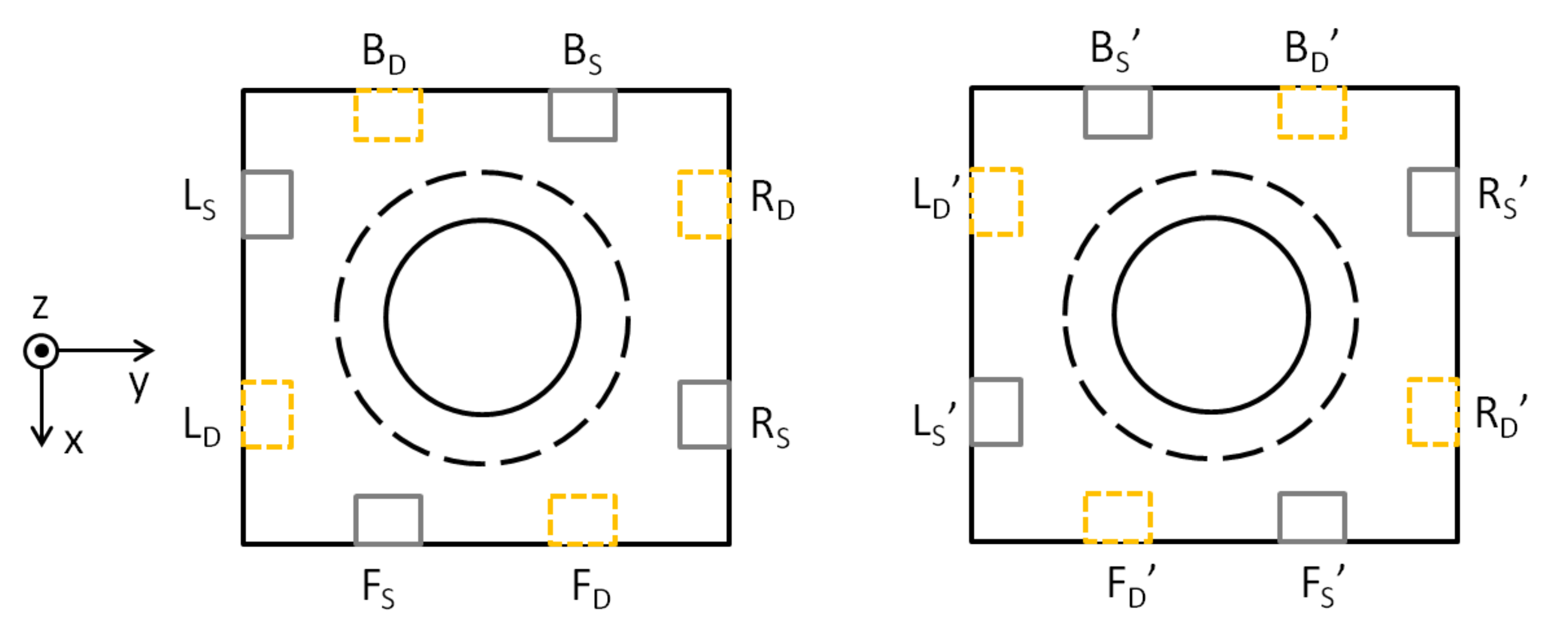}
  \caption{Top (left figure) and bottom (right figure) section of apparatus at the light source and detector planes, x and y are the transverse direction, and z is the vertical direction. Detectors (in orange) are labeled with subscript D, light source with subscript S (in grey), with L, R, F and B indicating the locations. Note that the beams intersect the sphere at the $\pm45^{\circ}$ latitudes. The equator is marked as a dashed circle}
  \label{fig:3D_ill}
\end{figure}

The top and bottom layer light source and detector layouts are shown in Fig. \ref{fig:3D_ill}. Four differential signals labeled as (L)eft, (R)ight, (F)ront, and (B)ack are each obtained by routing a pair of two signals to a differential detector. There can be many ways for the selection of two signals as a differential pair. The configuration used in this work is shown in table \ref{tab:1}, where the relationship between the displacement $\Delta x$, $\Delta y$, $\Delta z$ and the four differential signals used is given. This grouping was chosen to simplify the layout of the optical fibers.

\begin{table}[!htb]
\caption{Left: Definition of differential measurement signals L, R, F and B used in our setup. 2nd row shows their relations to physical sensor units shown in Fig. \ref{fig:3D_ill}. For example, differential signal L is generated by subtracting $\rm{L_D}$ from $\rm{L_D'}$. 3rd row shows the corresponding displacement in $\Delta x$, $\Delta y$ and $\Delta z$. Right: Expression of displacements $\Delta x$, $\Delta y$ and $\Delta z$ calculated from differential output signals L, R, F and B. Note that $\Delta z$ can also be obtained by (R+L)/2 in this example.}
  \hspace{0.3cm}
    \begin{minipage}{.5\linewidth}
      \centering
        \begin{tabular}{|c| c |c| c| c|}
        \hline
            Symbol  & L & R & F & B \\ \hline
            Definition & $\rm{L_D'-L_D}$ & $\rm{R_D-R_D'}$ & $\rm{F_D-F_D'}$ & $\rm{B_D'-B_D}$ \\ \hline
            Expression & $\Delta x+\Delta z$ & $-\Delta x+\Delta z$ & $-\Delta y-\Delta z$ & $-\Delta y+\Delta z$ \\ \hline
        \end{tabular}
    \end{minipage}%
    \hspace{1.0cm}
    \begin{minipage}{.5\linewidth}
      \centering
        \begin{tabular}{|m{0.5cm}|m{0.6cm}|m{0.6cm}|@{}m{0pt}@{}}
        \hline
            $\Delta x$ & $\Delta y$ & $\Delta z$ \hfill \\ \hline
            $\frac{\rm{L-R}}{2}$ & $\frac{\rm{-F-B}}{2}$ & $\frac{\rm{-F+B}}{2}$&\\[7pt]  \hline
        \end{tabular}
    \end{minipage}

\label{tab:1}
\end{table}



\begin{figure}[htbp]
  \centering
  \vspace*{-0.3cm}
  \includegraphics[width=0.9\textwidth]{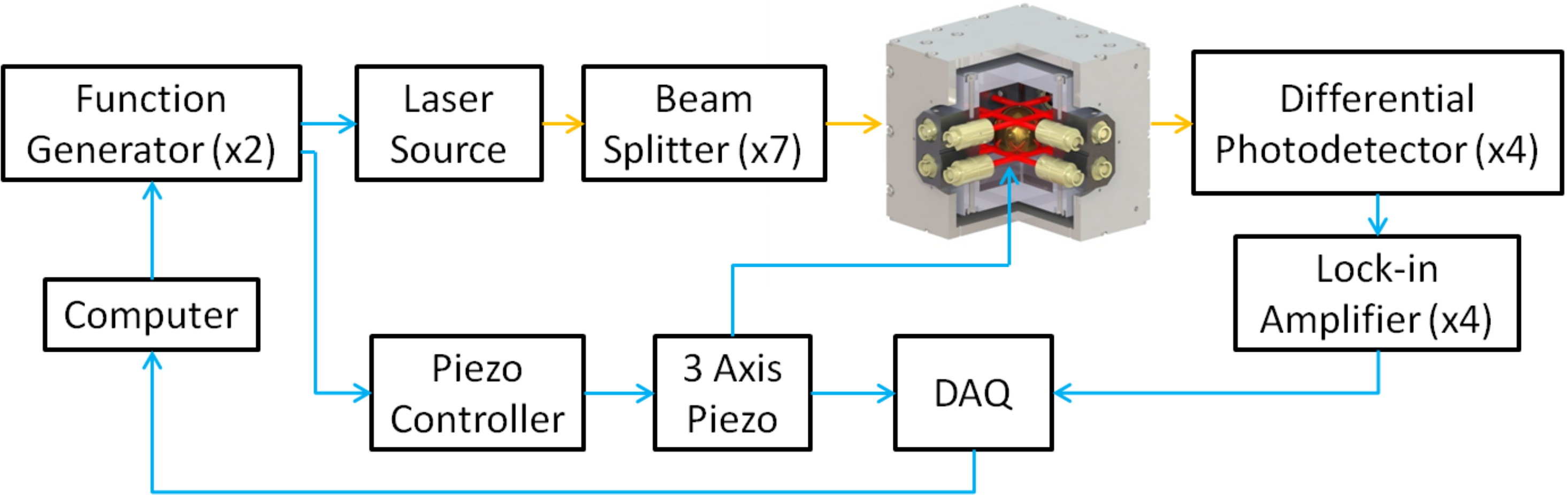}\vfill
  \caption{DOSS system diagram. Blue: electric signal. Orange: optical signal. DAQ: Data acquisition system}
  \label{fig:DOSS_sys}
\end{figure}

The DOSS system diagram for the experiment is shown in Fig. \ref{fig:DOSS_sys}. For 1D/3D measurements, the spherical TM is driven by the 3-axis piezo actuator shown in Fig. \ref{fig:DOSS_setup} in the specific direction(s) of interest, which is controlled through the piezo controller. For each axis, a 2-channel function generator is used to modulate the laser intensity for lock-in detection at 1.36 kHz, and to drive the piezo actuator (and TM) with specific patterns. We measured that the use of Lock-in amplifiers reduces the noise by $\sim 8$ dB. The laser source is split into 8 beams, and intensity pairs are measured by 4 differential photodetectors that are then demodulated in lock-in amplifiers, the outputs of which are recorded by DAQ system controlled by a desktop computer. 

\section{Noise analysis}\label{theory}
\subsection{Thermal noise}
\begin{figure}[htbp!]
  \centering
  \vspace*{-0.5cm}
  \includegraphics[width=0.6\textwidth]{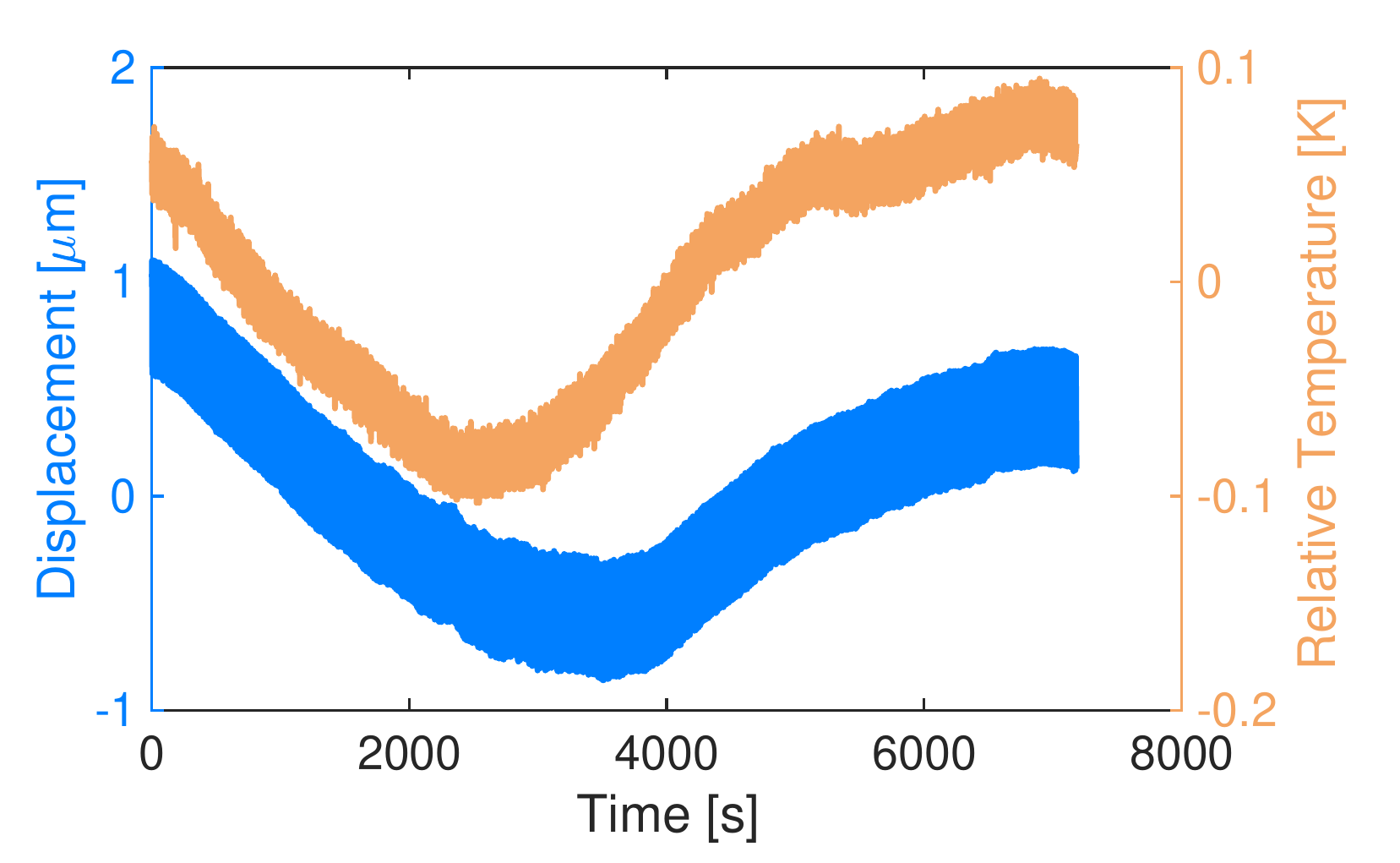}
  \caption{ Measured room temperature and displacement profile in time domain.}
  \label{fig:thermal}
\end{figure}

Thermal effects from various mechanisms can induce force/displacement between TM and housing, and their contribution can be a limiting factor for the lower end of the DOSS noise spectrum. In our current test setup, thermal control and vacuum environment are not implemented and the TM is rigidly mounted, hence unbalanced thermal expansion is likely to be the dominant effect. We compared the outer housing temperature with one channel of the differential DOSS readout, obtaining the results shown in Fig. \ref{fig:thermal}. It can be seen that the DOSS displacement measurement shows some correlation with environmental thermal variation. A time lag of $\sim$900 s between temperature variation and DOSS readout is observed We estimate the temperature-displacement coupling coefficient to be $\sim7.8\:\rm{\mu m/K}$. Of course temperature variations inside the housing will also affect the readout with even shorter time constant, which will give rise to noise in the displacement spectrum above 1 mHz.

In a space environment four types of thermal noise can be expected: the radiometer effect ($\propto P\cdot \delta T/T$), the residual gas effect ($P^{1/2}T^{1/4}$), the radiation pressure asymmetry effect ($\propto\delta T\cdot T^3$), and the temperature dependent asymmetric out-gassing of the housing ($\propto\delta T/T^2$); where $P$ is the residual gas pressure in the sensor housing and the acceleration noise from the four effects is inversely proportional to the mass of the TM \cite{schumaker2003disturbance}. For a drag-free system using the readout described in this paper and optimized for geodesy: $R_{\rm{TM}} = 1.27$ cm, gap to TM $\geq$ 1.23 cm, $M_{\rm{TM}} = 0.070$ kg, $T = 300$ K, $\delta T = 10^{-3} \rm{K\cdot{Hz}^{-1/2}}$ at 1 mHz rising as $(1/f)^{1/3}$, $P = 10^{-4}$ Pa, the main thermal disturbance is the radiometer effect $0.6\times10^{-12}\:\rm{m\cdot s^{-2}Hz^{-1/2}}$ \cite{stanford_report}. The other effects are more than an order of magnitude smaller. A design optimized for LISA  would show much better performance.


\subsection{Shot noise}
The shot noise of a photodetector represents the minimum detectable signal under ideal conditions. It can be expressed as:
\begin{equation}
S_{\rm{p}}=\sqrt{2eI}=\sqrt{2{r^2}h\nu P/{\eta}}
\end{equation}
Where $e$ is the electron charge, $P\sim 15\mu W$ is the optical power, $h$ is the Planck's constant, $\nu$ is the photon frequency, $\eta\sim 1$ is the quantum efficiency of the photodetector, $r=\eta e/(h\nu)$ is the responsivity, $I=Pr$ is the electric current.

Assuming uniform optical power distribution across the beam, the conversion factor from $S_{\rm{p}}$ to displacement noise $S_{\rm{L,s}}$ in units of W/m is $D=P/d$, hence the shot noise expressed in displacement units is:
\begin{equation}
S_{\rm{L,s}}=\sqrt{\frac{2r^2h\nu} {P\eta}}\cdot d\quad\left[\rm{\frac{m}{\sqrt{Hz}}}\right]
\label{eq:shot}
\end{equation}

Evaluating Eq. (\ref{eq:shot}) using experimental parameters, we obtain an estimated shot noise limit for DOSS of $S_{\rm{L,s}}\approx 0.29 \rm{nm/\sqrt{Hz}}$.

\subsection{Other noise sources}
\subsubsection*{Seismic and acoustic noise}
Seismic and acoustic noise can be coupled into the TM through the piezo actuated stage to which it is attached. It was measured in a similar lab environment to be $\sim 3 \rm{nm/\sqrt{Hz}}$ at 1 Hz along the vertical axis \cite{tan2017pico}. Since only the differential residual (TM+piezo actuator and TM housing are common mode) of the seismic/acoustic noise will affect the measurement, we expect this to have negligible effect for the transverse directions. The seismic induced effect is expected to appear in the measurement of piezo actuator and electronics noise, as introduced in the next section.

\subsubsection*{Piezo actuator and electronics noise}
\begin{figure}[htbp!]
  \centering
  \includegraphics[width=0.6\textwidth]{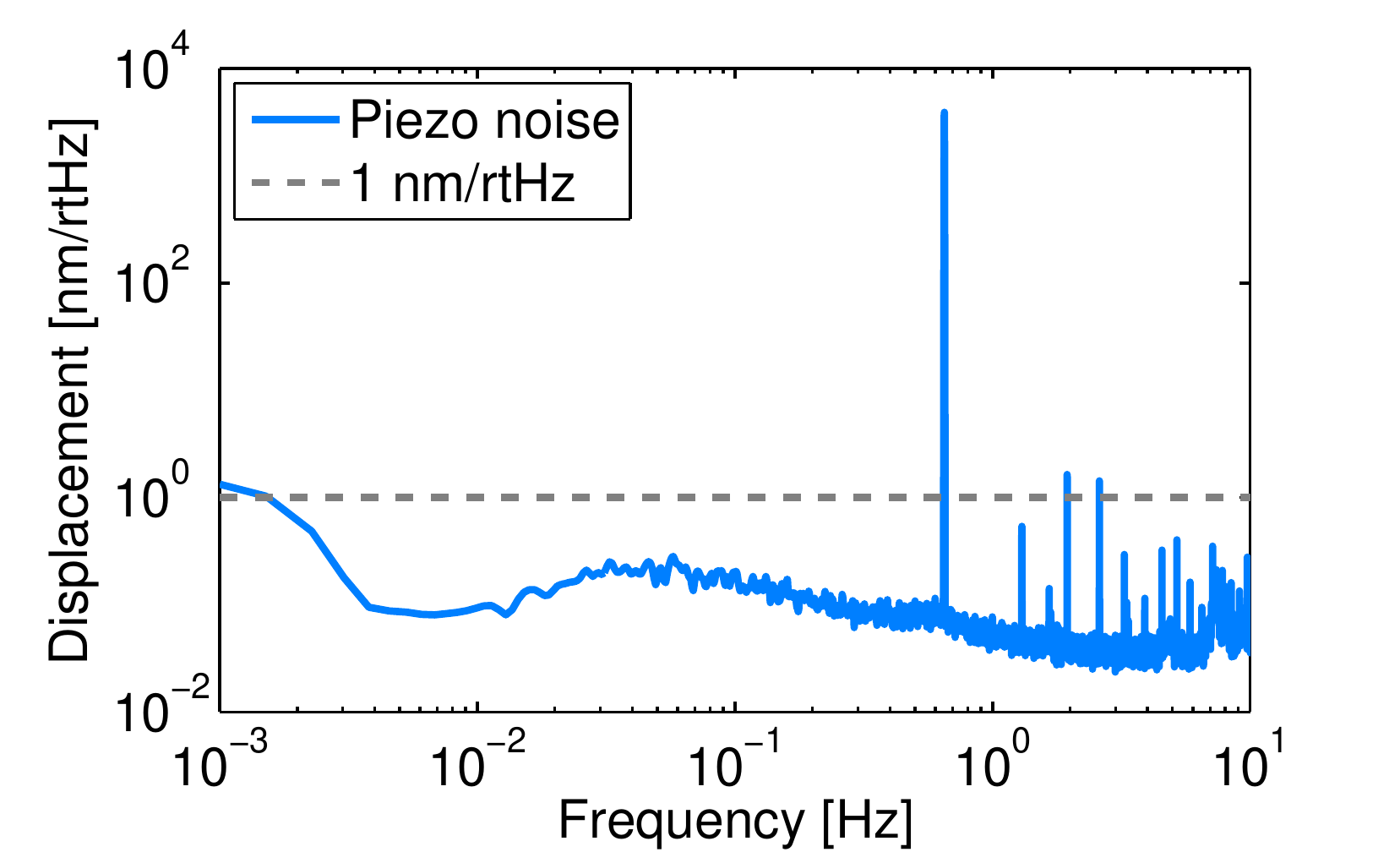}
  \caption{Single axis piezo actuator noise spectrum. Spike at 0.65 Hz is a calibration signal.}
  \label{fig:piezo}
\end{figure}

The noise contribution of piezo actuator, corresponding driver electronics and DAQ , with the presence of seismic and acoustic noise, was measured, by displacing the piezo actuators by a known amount and recording the piezo built-in displacement output through the DAQ system. The resulting spectrum is shown in Fig. \ref{fig:piezo}.

The spike at 0.65 Hz is due to the calibration signal, and as can be seen, the noise contribution from the actuator system is well below the expected DOSS displacement noise $\sim1\:\rm{nm/\sqrt{Hz}}$ so it does not contribute much to the overall DOSS noise measurement.


\section{Results}
\label{results}
\subsection{1D measurements}
\begin{figure}[htbp!]
  \centering
  \includegraphics[width=0.6\textwidth]{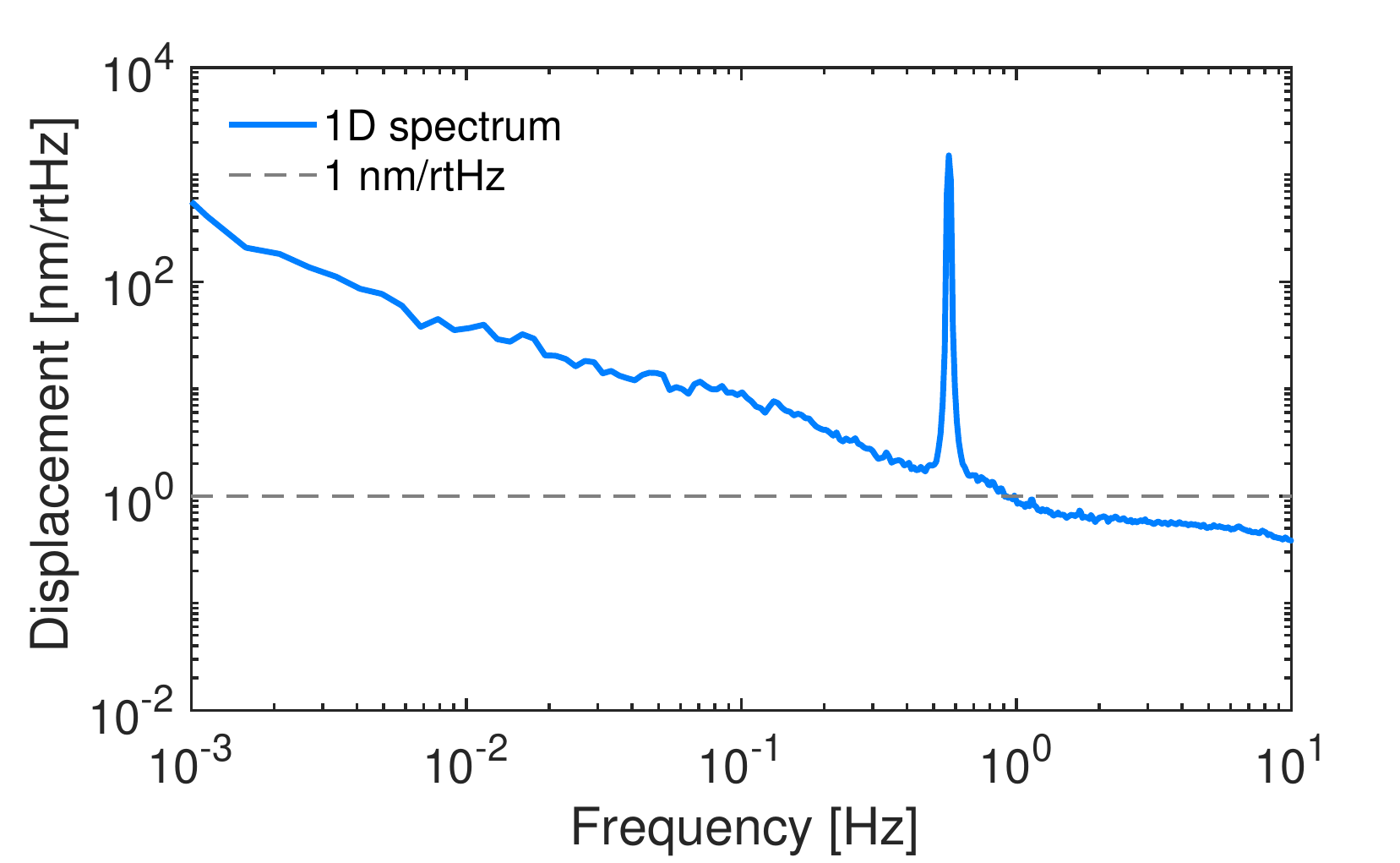}
  \caption{ASD of 1D DOSS displacement noise measured in the Y-direction.}
  \label{fig:1D}
\end{figure}

The results obtained by displacing the TM in one direction are shown in Fig. \ref{fig:1D}. As stated in section \ref{setup}, the TM is displaced by a specified amount and pattern, using the piezo actuator. The actuator has a built-in displacement measurement device that can be used as a calibration signal for the DOSS measurement. In this case, the calibration signal is a sine wave with a peak-peak amplitude of 507 nm at 0.65 Hz, which gives rise to the width of the displacement data band in Fig. \ref{fig:thermal} and leads to the corresponding peak in the ASD plot in Fig. \ref{fig:1D}.

As can be seen from Fig. \ref{fig:1D}, 
the displacement measurement shows an increase of noise for frequencies $<$1 Hz. We suspect that this is partially due to thermal and environmental fluctuations throughout the system. In future work, we anticipate placing the DOSS unit in a thermal controlled multi-layer structure \cite{alfauwaz2016thermal} with a temperature stability of $<100$ \textmu K, where thermally induced perturbations can be greatly reduced.

The displacement spectral noise was measured to be $0.87\:\rm{nm/\sqrt{Hz}}$ at 1 Hz , and $0.39\:\rm{nm/\sqrt{Hz}}$ at 10 Hz. Below 1 Hz, long term effects including thermal, seismic and various other technical noise sources are expected to dominate, leading to a noise amplitude of $\sim400\:\rm{nm/\sqrt{Hz}}$ at 1 mHz.


\subsection{3D measurements}
\begin{figure}[htbp!]
	  \centering
	  \includegraphics[scale=0.24]{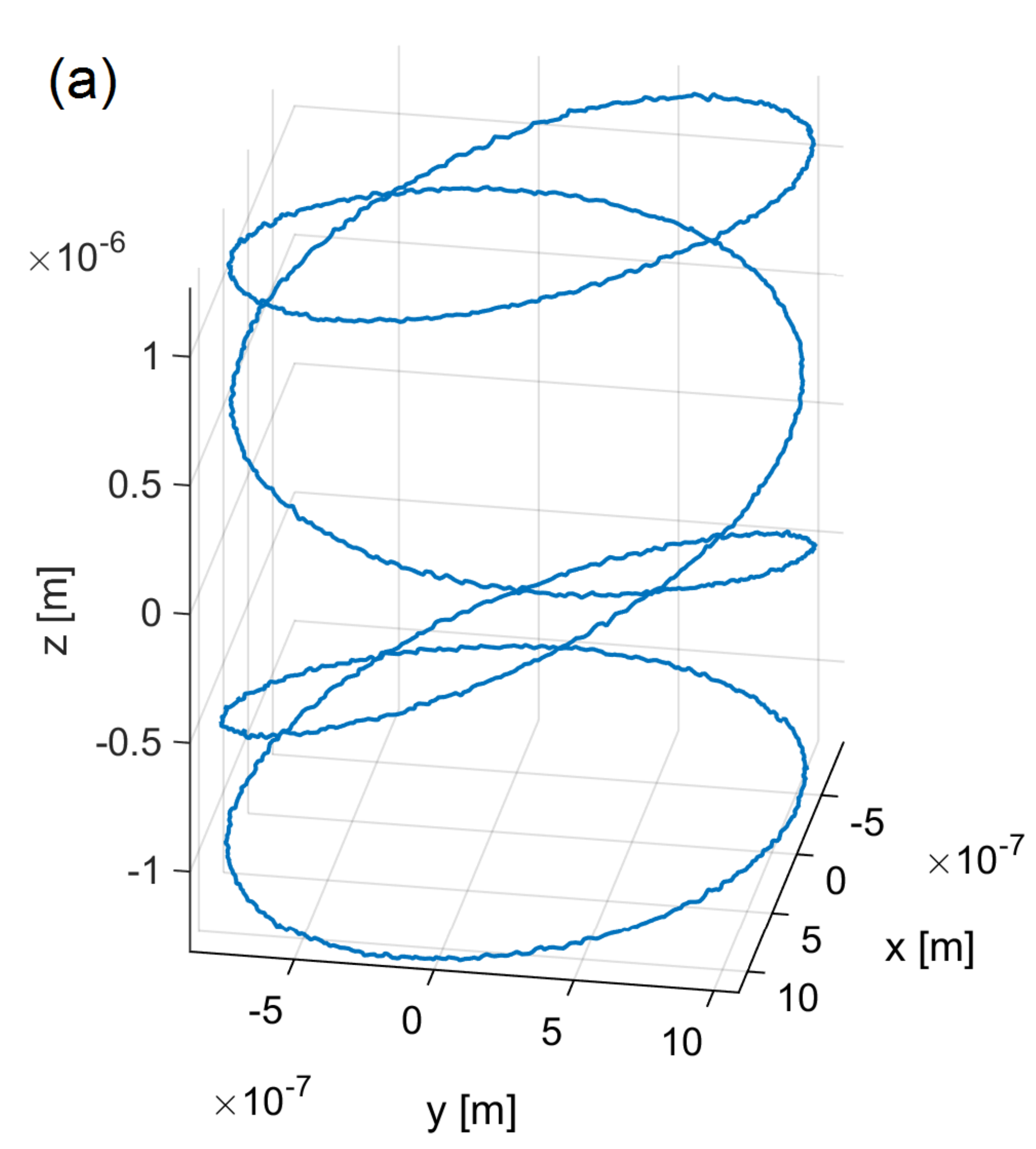}~
	  \includegraphics[scale=0.53]{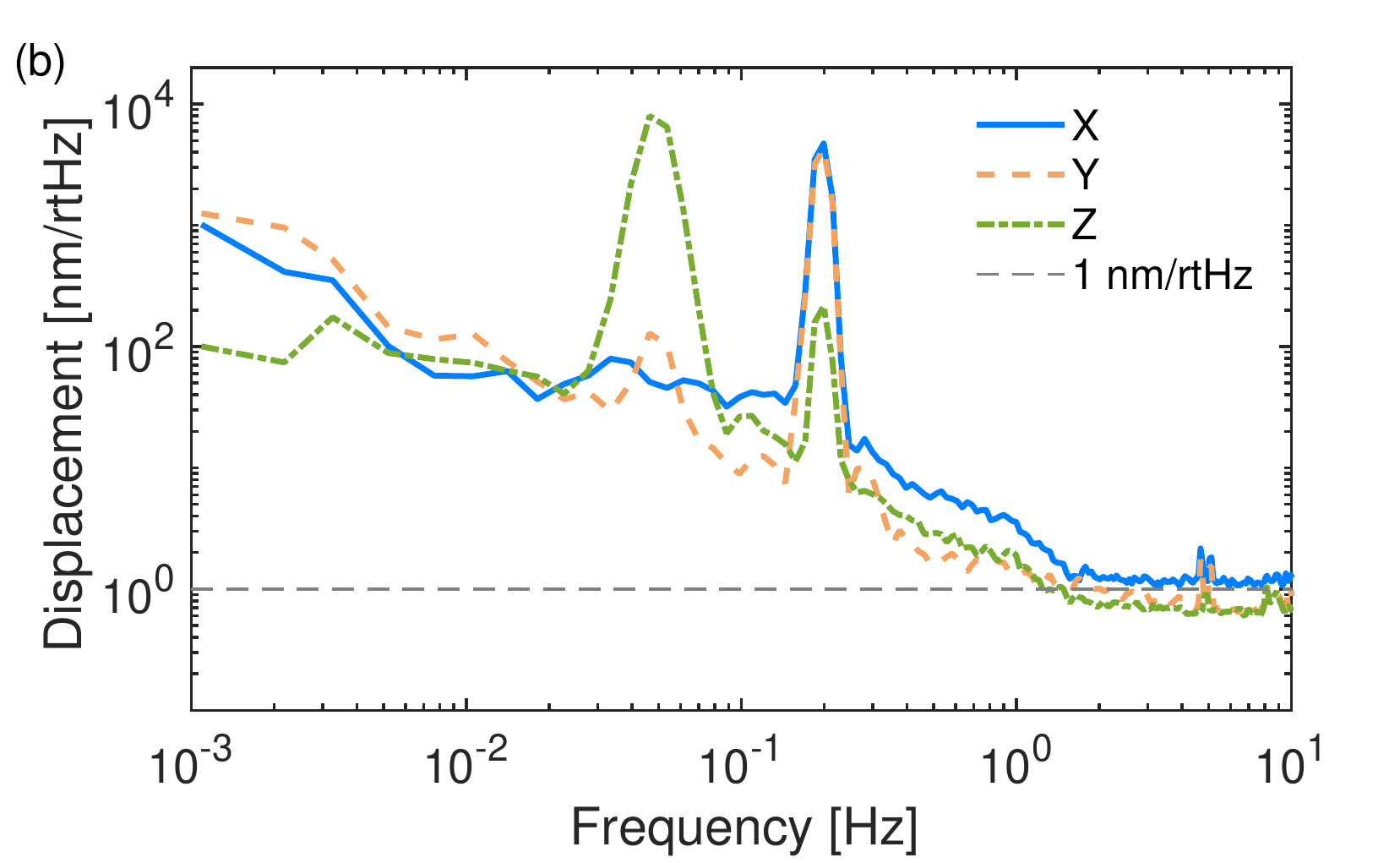}
	  \caption{3D DOSS measurement results. (a). Measurement trajectory. (b). Displacement ASD}
	  \label{fig:3D}
\end{figure}

As in the 1D case, for the 3D measurement we drive the TM using piezos in preset patterns, and three sets of DOSS data are measured using four differential laser/photodiode pairs as indicated in Table \ref{tab:1}. 
The trajectory of the 3D displacement is shown in Fig. \ref{fig:3D}(a), where we displace the TM in a circle in the XY plane at 200 mHz for both directions, and in the Z direction the TM is moved in a sine pattern at 50 mHz. The total travel range for X, Y and Z directions are 1.9 \textmu m, 1.9 \textmu m and 2.5 \textmu m, respectively.

The ASDs for the three axes are shown in Fig. \ref{fig:3D}(b). It can be seen that the 3D ASDs follow a similar trend to the 1D case. The spectral noise for the X direction appears to be somewhat higher than Y and Z. We note that cross-coupling between the measurement channels can be seen in the figure. For example, the Z profile shows peaks at both drive frequencies. In the ideal case, if we actuate in the X direction alone, with the absence of cross-coupling, only R and L channels should give non-zero output according to Table. \ref{tab:1}. However, with its presence, we measure a -37 dB cross-couping signal from X to channels F and B with a spectrum analyzer. Similarly we can do this for the Y axis, which shows a -39 dB coupling to channel L, and a -41 dB coupling to R. For comparison with other devices, we note the the overall resolution for the 3D displacement measurement is $\sim1\:\rm{nm/\sqrt{Hz}}$ at 1 Hz. At very low frequencies we find the noise spectrum to be somewhat variable, indicating it is dominated by environmental effects. This situation would be much more benign in a LISA-like observatory due to the choice of orbit.


\section{Conclusion}
\label{conclusion}
We have demonstrated a method of high sensitivity displacement measurement based on a differential optical shadow sensor (DOSS), in a configuration suitable for use in a drag free satellite. The measurement scheme allows for large gaps between the TM and the housing, which reduce the perturbations to the TM. Other advantages of the DOSS sensor for drag-free applications are its simplicity and low volume/mass budget. This makes it feasible to integrate a GRS based on the DOSS into small satellites. A technology demonstration mission for the DOSS sensor was developed \cite{Zoellner2012Differential} and similar sensor was designed for a drag-free CubeSat \cite{Conklin2012DragFree}. 

We have demonstrated the proof of concept of a sensor capable of simultaneous 3D displacement measurements of a TM, with a sensitivity better than $1\:\rm{nm/\sqrt{Hz}}$ at 1 Hz. It is feasible to use it as stand-alone displacement sensor for drag-free applications such as geodesy. A number of missions \cite{tapley2004gravity,Canuto2008DragFree} have already mapped the Earth's gravity field with sufficient precision to detect the effects of droughts and mass loss of ice sheets. The device described here can be used to obtain improved measurements from future missions with only minor upgrades for launch qualification. For missions with stricter requirements such as gravitational wave detection, the DOSS can be used as sensor for the drag-free control in combination with an additional interferometric sensor for the scientific measurement, as was the case in the LISA Pathfinder demonstration mission. In addition, it is also possible to be used for ground based applications such as accelerometers and roundness measuring devices. In future work we hope to reduce the low frequency noise that appears to be due to environmental effects, as suggested by the data in Fig. \ref{fig:thermal}.

\section*{Funding}
King Abdulaziz City for Science and Technology (KACST) funding $\#$1169603.

\section*{Acknowledgments}
We thank R. Byer (Stanford University) and Dr T. Al Saud (KACST) for supporting the project. 

\end{document}